\documentclass[preprint,showpacs,preprintnumbers,amsmath,amssymb]{revtex4}
\usepackage{graphicx}
\usepackage{floatflt}
\usepackage{dcolumn}
\usepackage{bm}
\usepackage{rotating}
\usepackage{natbib}
\usepackage{txfonts}

\begin{document}
\title{  Stable configurations of hybrid stars with colour-flavour-locked core}

\author{B. K. Agrawal$^{1}$}\email{bijay.agrawal@saha.ac.in}
   \author{Shashi K. Dhiman$^{2,3}$}\email{shashi.dhiman@gmail.com}  
     \affiliation{
$^{(1)}$Saha Institute of Nuclear Physics, Kolkata - 700064, India.\\
$^{(2)}$Department of Physics, Himachal Pradesh University, Shimla - 171005, India.\\
$^{(3)}$University Institute of Information Technology, Himachal Pradesh University, Shimla 171005, India.} 


\begin{abstract}

We construct static and mass-shedding limit sequences of hybrid stars,
composed of colour flavour locked (CFL) quark matter core, for a set
of equations of state (EOSs).  The EOS for the hadronic matter is
obtained  using appropriately calibrated extended field theoretical
based relativistic mean-field model.  The MIT bag model is employed to
compute the EOSs of the CFL quark matter for different
values of the CFL gap parameter in the range of $50 - 150\text{ MeV}$ with the
deconfinement phase transition density ranging from $4\rho_0 - 6\rho_0$
($\rho_0 = 0.16\text{ fm}^{-3}$).  We find,  depending on the values of the
CFL gap parameter and the deconfinement phase transition density, 
the sequences of stable configurations of hybrid stars either form third
families of the compact stars or bifurcate from the hadronic sequence.
The hybrid stars have masses $1.0 - 2.1 M_\odot$ with radii $9 - 13.5$ km.
The maximum values of mass shedding limit frequency for such hybrid stars
are $1 -2$ kHz.  For the smaller values of the  CFL gap parameter and  the
deconfinement phase transition density, mass-radius relationships are in
harmony with those deduced by applying improved hydrogen atmosphere model
to fit the high quality spectra from compact star X7 in the globular
cluster 47 Tucanae.  We observed for some cases that the third family
of compact stars exist in the static sequence, but, disappear from the
mass-shedding limit sequence.  Our investigation suggests that
the third family of compact stars in the mass-shedding limit sequence
is more likely to appear, provided they have maximum mass in the static
limit  higher than their second family counterpart composed of pure
hadronic matter.

\end{abstract}

\maketitle

\section{Introduction}
\label{intro_sec}

Soon after the suggestion that three-flavor quark matter may be the
ground state of strongly interacting systems \cite{Witten84}, quark
stars are postulated as possible astrophysical objects. It was also
hypothesized that some compact stars might be  hybrid stars with
the core composed of quark matter and surrounded by nuclear mantle.
The present knowledge of quantum chromodynamics (QCD) at high density
indicates that quark matter might be in a color superconducting
phases.  The essence of color superconductivity is quark-quark color
superconductor \cite{Bailin84,Alford01} and driven by Bardeen, Cooper
and Schrieffer (BCS) \cite{Bardeen57,Bardeen57a} pairing mechanism.
The possible quarks color superconducting phases include the two-flavor
color superconductor (2SC) \cite{Alford98a,Ruster04,Ruster05}, the
colour flavour locked (CFL) phase \cite{Alford99a,Rajagopal01}, and the
crystalline color superconductor (CCS)
\cite{Alford01a,Rajagopal06,Ippolito07}.  The speculation that the
colour superconducting quark matter present in the core of the hybrid
stars has triggered many theoretical investigations.

The hybrid stars  with CFL quark matter core have been extensively
studied. The hadron phase of the hybrid star matter is described by the
various models which can be broadly grouped into (i) non-relativistic
potential models \cite{Pandharipande75}, (ii) non-relativistic
mean-field models \cite{Chabanat97,Stone03,Mornas05,Agrawal06},
(iii) field theoretical based relativistic mean-field
models (FTRMF) \cite{Prakash95,Glendenning99,Steiner05}
and (iv) Dirac-Brueckner-Hartree-Fock model
\cite{Muther87,Engvik94,Engvik96,Schulze06}.   The CFL quark matter
appearing at the core of hybrid stars are described within the MIT bag
model and Nambu-Jona-Lasinio (NJL) model.  The  studies based on the MIT bag model indicate
the existence of  stable configurations of hybrid stars with the CFL
quark matter core \cite{Alford03,Banik03,Alford04}. Two different
situations are encountered, the hybrid stars with CFL quark matter
core  either form a third family of compact stars separated from the
purely hadronic sequence by an instability region or bifurcate from
the hadronic sequence of stars when the central density exceeds the
phase transition density  at which deconfinement of hadrons to the CFL
quark matter occurs.  The scenario is completely different when NJL model
is employed to study the hybrid stars with CFL quark matter core. Until
recently \cite{Baldo03,Buballa05,Klahn07}, it was shown that the NJL
like model rules out  the CFL quark matter phase at the core because
it renders the hybrid star unstable. Only very recently, it has been
found that large enough values of the diquark coupling strength in NJL
model can yield  stable configurations of the hybrid star containing
CFL quark matter core \cite{Ippolito08,Pagliara08}.

The stability of hybrid star  with CFL quark matter core depends strongly
on the values of the deconfinement phase transition density  and the CFL
gap parameter which are poorly known.  In the present work we construct the
static sequences of hybrid stars, with CFL quark matter
core, for a set of EOSs obtained for different values of CFL gap parameter
and the deconfinement phase transition density.  The hadron phase of
the hybrid star is described by using an appropriately calibrated
extended FTRMF model which includes the contributions from self- and
mixed interaction terms for $\sigma$, $\omega$ and $\rho$ mesons up to
the quartic order.  The CFL quark matter phase is described within the
MIT bag model with an additional parameter that mimics the effect of
including perturbative QCD corrections.  Instead of keeping  the value
of the bag constant fixed as previously done \cite{Alford03,Banik03},
calculations are performed for different values of the CFL gap parameter
at fixed values of the deconfinement phase transition density.  This
strategy should enable us to asses better the influence of the CFL gap
parameter on the properties of hybrid stars with CFL quark matter  core.
The CFL gap parameter $\Delta$ is varied in the range of $50 -
150\text{ MeV}$ by keeping the deconfinement phase transition density $\rho_t$
fixed in between $4\rho_0 - 6\rho_0$ ($\rho_0 = 0.16\text{fm}^{-3}$).
For  the different values of the CFL gap parameter considered, the
average quark chemical potential at the deconfinement phase transition
density lie in the range of $375 - 500\text{ MeV}$ which is in reasonable
agreement with the predictions of the NJL model.

The paper is organized as follows, in Sec. II we describe  in brief the
models used to construct the EOSs for hadronic phase, CFL quark phase,
and the mixed phase. In Sec. III we present  the results for static
and mass-shedding limit sequences for hybrid stars.  In Sec. V we state
our conclusions.

\section { Equations of State for Hybrid Star matter}
\label{sec:model1}
 
We construct the EOS for the hybrid star matter which is composed
of hadrons at low densities, quark matter in the CFL phase at high
densities and the mixed phase at the intermediate densities. The EOS
for the  hadron matter is obtained within the framework of the extended
FTRMF model. The EOS for the quark matter in the CFL phase is obtained using
the MIT bag model. The EOS for the mixed phase is constructed using the
Gibbs conditions.  For the hadron matter at very low densities $\rho \sim
0.5\rho_0\text{ fm}^{-3}$ going down to  $\rho = 6.0\times 10^{-12}\text{ fm}^{-3}$
we use Negele-Vautherin \cite{Negele73} and Baym-Pethick-Sutherland
EOS \cite{Baym71}.

\subsection{Hadron phase}

The hadronic phase is described using the extended FTRMF model which
includes the contributions from self- and mixed interaction terms for
$\sigma$, $\omega$, and $\rho$ mesons up to the quartic order.  The mixed
interaction terms involving the $\rho$-meson field enables one to vary
the density dependence of the symmetry energy coefficient and neutron skin
thickness in heavy nuclei over the wide range without affecting the other
properties of the finite nuclei \cite{Furnstahl02,Sil05}. The contribution
from the self-interaction of $\omega$-mesons plays an important role in
determining the high density behavior of EOS and consequently the
structure properties of compact stars \cite{Dhiman07,Muller96}.
The contributions of self-interaction of $\rho$-meson are ignored as
they  affect the ground state properties of heavy nuclei and compact
stars only very marginally \cite{Muller96}.  In our recent work
\cite{Dhiman07} we have obtained several parameterizations of the
extended FTRMF model in such a way that the bulk nuclear observables
and nuclear matter incompressibility coefficient are fitted well.
These different parameterizations produce different behavior for the
EOS at high densities.

The energy density of the hadron phase in the extended FTRMF models is given by
 \begin{equation}
\label{eq:eden}
\begin{split}
{\cal E}_{HP}  (\mu_n,\mu_e)&= \frac{1}{\pi^{2}}\sum_{j=n,p}\int_{0}^{k_f^j}k^2\sqrt{k^2+M^{*2}} dk
+g_{\omega N} \omega(\rho_p+\rho_n)+\frac{1}{2}g_{\rho N}{\rho}(\rho_p -\rho_n)
+ \frac{1}{2}m_{\sigma}^2\sigma^2\\
&+\frac{\overline{\kappa}}{6}g_{\sigma N}^3\sigma^3
+\frac{\overline{\lambda}}{24}g_{\sigma N}^4\sigma^4
-\frac{\zeta}{24}g_{\omega N}^4\omega^4
 - \frac{1}{2} m_{\omega}^2 \omega ^2
-\frac{1}{2} m_{\rho}^2 \tilde\rho ^2\\
&-\overline{\alpha_1} g_{\sigma N}
 g_{\omega N}^{2}\sigma \omega^2-\frac{1}{2} 
\overline{\alpha_1}^\prime g_{\sigma N}^2 g_{\omega N}^2\sigma^2 \omega^2
-\overline{\alpha_2}g_{\sigma N}g_{\rho N}^2 \sigma\rho^2
 -\frac{1}{2} \overline{\alpha_2}^\prime g_{\sigma N}^2 g_{\rho N}^2\sigma^2 
\rho^2\\
& - \frac{1}{2} \overline{\alpha_3}^\prime g_{\omega N}^2 g_{\rho N}^2
\omega^2\rho^2+\frac{1}{\pi^{2}}\sum_{l=e^-,\mu^-}\int_{0}^{k_f^l}k^2\sqrt{k^2+m^2_l} dk 
\end{split}
\end{equation}
The pressure of the hadron phase matter  is given by
\begin{equation}
\label{eq:press}
\begin{split}
P_{HP} (\mu_n,\mu_e)&= \frac{1}{3\pi^{2}}\sum_{j=n,p}\int_{0}^{k_f^j}
\frac{k^{4}dk}{\sqrt{k^2+M^{*2}}} 
- \frac{1}{2}m_{\sigma}^2\sigma^2-\frac{\overline{\kappa}}{6}g_{\sigma N}^3\sigma^3 -\frac{\overline{\lambda}}{24}g_{\sigma N}^4\sigma^4\\
& +\frac{\zeta}{24}g_{\omega N}^4\omega^4
  + \frac{1}{2} m_{\omega}^2 \omega ^2
+\frac{1}{2} m_{\rho}^2 \rho ^2+\overline{\alpha_1} g_{\sigma N}
g_{\omega N}^{2}\sigma \omega^2\\
&+\frac{1}{2} \overline{\alpha_1}^\prime g_{\sigma N}^2 g_{\omega N}^2\sigma^2 \omega^2+\overline{\alpha_2}g_{\sigma N}g_{\rho N}^2 \sigma\rho^2
 +\frac{1}{2} \overline{\alpha_2}^\prime g_{\sigma N}^2 g_{\rho N}^2\sigma^2 
\rho^2\\ 
& + \frac{1}{2} \overline{\alpha_3}^\prime g_{\omega N}^2 g_{\rho N}^2
\omega^2\rho^2+\frac{1}{3\pi^{2}}\sum_{l=e^-,\mu^-}\int_{0}^{k_f^l}\frac{k^{4}dk}{\sqrt{k^2+m^2_l}}
\end{split}
\end{equation}
where, $M^{*} = M - g_{\sigma N }\sigma $ is the effective mass of
nucleon with $M$ being the free nucleon mass.  In Eqs. (\ref{eq:eden}) and (\ref{eq:press}), $\sigma$, $\omega$
and $\rho$ represent the meson fields.  The $g_{\sigma N}$, $g_{\omega
N}$ and $g_{\rho N}$ are the meson-nucleon coupling strengths. The $m_\sigma$,
$m_\omega$ and $m_\rho$ are the  masses for the $\sigma$, $\omega$ and
$\rho$ mesons.  The coupling strengths for the self-interaction terms
for the $\sigma$ and $\omega$ mesons are denoted by $\overline{k}$,
$\overline{\lambda}$ and $\zeta$. The constants $\overline{\alpha}$
and $\overline{\alpha}^\prime$ represent the coupling strengths for
various mixed interaction terms. The last term in Eqs. (\ref{eq:eden})
and (\ref{eq:press}) gives the contributions to the energy density and
pressure from leptons,  respectively.
In Eqs. \ref{eq:eden} and \ref{eq:press}, $\mu_n$ and $\mu_e$ represent the
chemical potentials for the neutrons and electrons, respectively. The
chemical potentials for the protons $\mu_p$  and the muons $\mu_\mu$ can be
expressed in terms of $\mu_n$ and $\mu_e$  using the
$\beta-$equilibrium conditions, i.e., 
\begin{eqnarray}
\mu_n = \mu_p + \mu_e\\
\mu_e = \mu_\mu.
\label{eq:beta_eq}
\end{eqnarray}
Once th chemical potentials for the  nucleons are known, their
fermi-momenta can be obtained by solving the field equations for the
mesons as given in Ref. \cite{Dhiman07}. The fermi-momenta  $k^l_f$ for the leptons
are obtained as,
\begin{equation}
k^l_f=\sqrt{\mu_l^2-m_l^2}.
\label{eq:fm_lep}
\end{equation}   In addition to the conditions of the $\beta-$equilibrium,
matter in pure hadronic phase is considered to be charge neutral.

\subsection{Quark matter in the CFL phase}
The free energy density for quark matter in the CFL phase is taken to be
\cite{Alford05},
\begin{equation}
\Omega_{CFL}(\mu,\mu_e)=\Omega_{CFL}^{quarks}(\mu)+\Omega_{CFL}^{GB}(\mu,\mu_{e})+\Omega^{electron}(\mu_{e}).
\label{eq:omega_cfl}
\end{equation}
where, $\mu$ is the average chemical potential for quarks and
$\mu_e$ is the electron  chemical potential.  The contribution to
Eq. (\ref{eq:omega_cfl}) from the quarks is given by,
\begin{equation}
\label{eq:omega_quarks}
\Omega_{CFL}^{quarks}= \frac{6}{\pi^2} \int_0^\nu p^2 (p-\mu)dp
+\frac{3}{\pi^2} \times \int_0^\nu p^2(\sqrt{p^2+m_s^2}-\mu)dp
+\frac{3}{4\pi^2} c\mu^4 - \frac
{3\Delta^2\mu^2}{\pi^2 }+B
\end{equation}
where, $u$ and $d$ quarks are assumed to be massless and $s$ quark has the
mass $m_s$.
The term proportional to $\mu^4$ in Eq. (\ref{eq:omega_quarks})
corresponds to the QCD inspired correction \cite{Fraga01}.
The second last term involving the CFL gap parameter $\Delta$ is the lowest
order contribution from the formation of the CFL condensate. 
The last term $B$ is  the bag model constant
which accounts for the energy difference between the perturbative vacuum and
the true vacuum.
The number densities for all the three flavours of
quarks considered are the same and can be obtained as, 
\begin{equation}
\label{eq:rhoq}
\rho_q=\frac{1}{\pi^2}(\nu^3 - c \mu^3+2\Delta^2\mu)
\end{equation}
with $q = u, d $ and $s$ and the common fermi momentum $\nu$ given as,
\begin{equation}
\nu=2\mu - \sqrt{\mu^2+\frac{m_s^2}{3}}.
\end{equation}
The contribution to the Eq. (\ref{eq:omega_cfl}) from the Goldstone bosons
arising due to the breaking of chiral symmetry in the CFL phase is
evaluated using $\Omega_{CFL}^{GB}(\mu,\mu_e)$ \cite{Son00}, 
\begin{equation}
\Omega_{CFL}^{GB}(\mu,\mu_{e})=-
\frac{1}{2}f^2_{\pi}\mu_e^2 \left( 1-\frac{m^2_{{\pi}^-}}{\mu_e^2}\right)^2
\end{equation}
where the parameters  are 
 \begin{equation}
f^2_{\pi}= \frac{(21-8 {\text ln} 2)\mu^2}{36\pi^2},
m^2_{{\pi}^-} = \frac{3\Delta^2}{\pi^2 f^2_{\pi}} m_s(m_u+m_d)
\end{equation}
\begin{equation}
\Omega^{lepton}(\mu_{e})=\frac{1}{\pi^2} \sum_{i=e^-,\mu^-}
\int_0^{\sqrt{\mu_e^2-m_i^2}} p^2(\sqrt{p^2+m_i^2}-\mu_e)dp.
\end{equation}
The total energy density and pressure for the quark phase (QP) can be
calculated as,

\begin{equation}
\label{eq:eden_cfl}
{\cal E}_{QP}=\Omega_{CFL}(\mu,\mu_{e})+3\mu \rho_{q}+\mu_e(\rho_e+\rho_{\mu}),
\end{equation}
and
\begin{equation}
P_{QP}=-\Omega_{CFL}(\mu,\mu_{e}),
\end{equation}
It is clear from the Eq. (\ref{eq:rhoq}) that the densities for the
$u$, $d$ and $s$ quarks are equal to each other   for the quark matter
in the CFL phase. Thus, quark matter in the CFL phase is enforced to
be  charge neutral. The electrons are present only in the mixed phase
of hadronic and the quark matter.

\subsection{Mixed phase}
The EOS for the mixed phase composed of hadronic and CFL quark matters is
obtained using the Gibbs conditions.  The Gibbs conditions can be expressed
in terms of two independent chemical potentials in our case as,
\begin{equation}
P_{HP}(\mu_n,\mu_e) = P_{QP}(\mu,\mu_e)
\end{equation}
where,  $\mu_n$ and $\mu_e$ are the two independent chemical potential
with $\mu_n$ being the neutron chemical potential.  In the mixed phase
the average quark chemical potential $\mu = \mu_n/3$. In the mixed
phase local charge neutrality condition is replaced by the global charge
neutrality 
 \begin{equation} \chi\rho^{ch}_{QP}+(1-\chi)\rho^{ch}_{HP}=0.
\label{eq:global_neut} \end{equation} 
 where $\chi$ is the volume
fraction occupied by quark matter in the mixed phase and $\rho^{ch}$
is the charge density.  It is clear from Eq. (\ref{eq:global_neut})
that both hadron and quark matter are allowed to be charged separately.
The energy density ${\cal E}_{MP}$ and the hadron  density $\rho_{MP}$
of the mixed phase can be calculated as \cite{Glendenning00},
 \begin{equation}
{\cal E}_{MP}=\chi{\cal E}_{QP}+(1-\chi){\cal E}_{HP}, \end{equation}
 \begin{equation}
\rho_{\scriptstyle MP}=\chi\rho_{\scriptstyle{QP}}+
(1-\chi)\rho_{\scriptstyle{HP}}, \end{equation} Once these quantities
are determined, we can construct the complete EOS with the hadron phase,
quark matter phase and  mixed phase and compute the properties of hybrid
compact star.

\section{Structure of hybrid stars with CFL core}

\begin{figure}
\includegraphics[width=14cm]{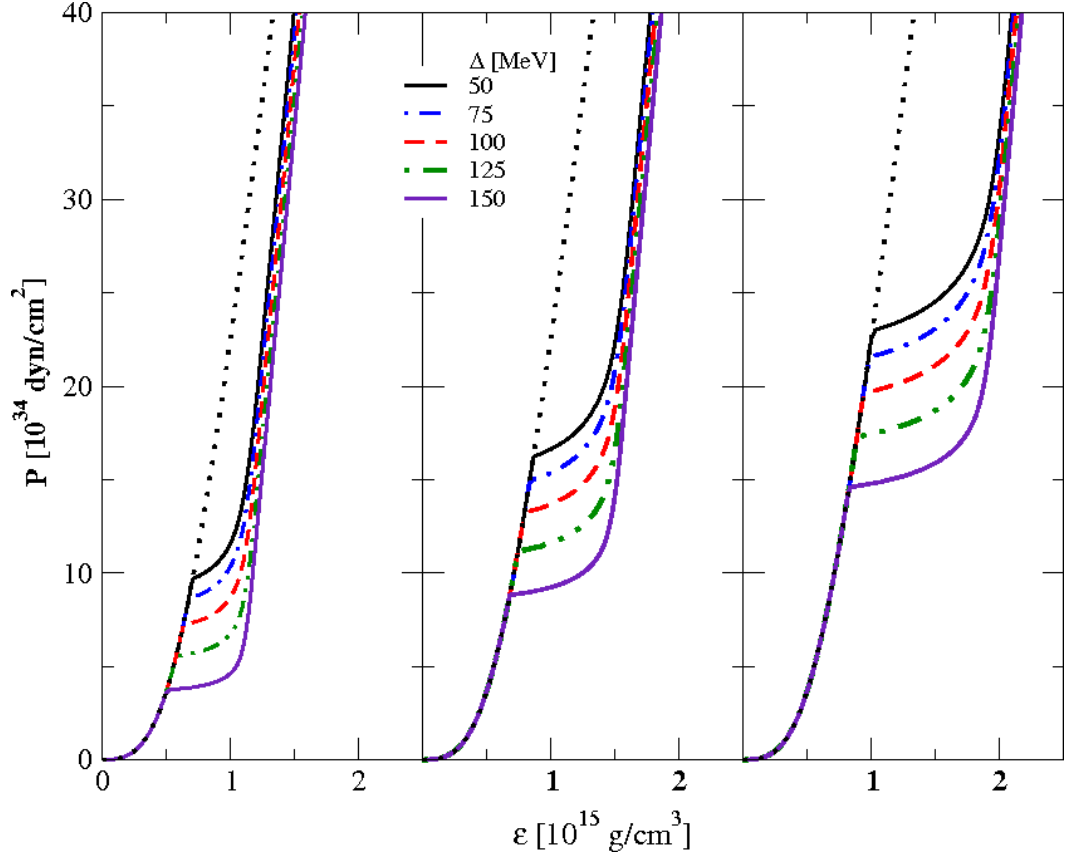}   
\caption{\label{fig:eos} (Color online)
Pressure as a function of energy density for different values of the CFL
gap parameter with hadron to CFL quark matter phase transition densities
$\rho_t = 4\rho_0$ (left), $5\rho_0$ (middle) and $6\rho_0$ (right).
The EOS for pure hadronic matter is shown by the black
dotted line.} 
 \end{figure}

\begin{figure}
\includegraphics[width=15cm]{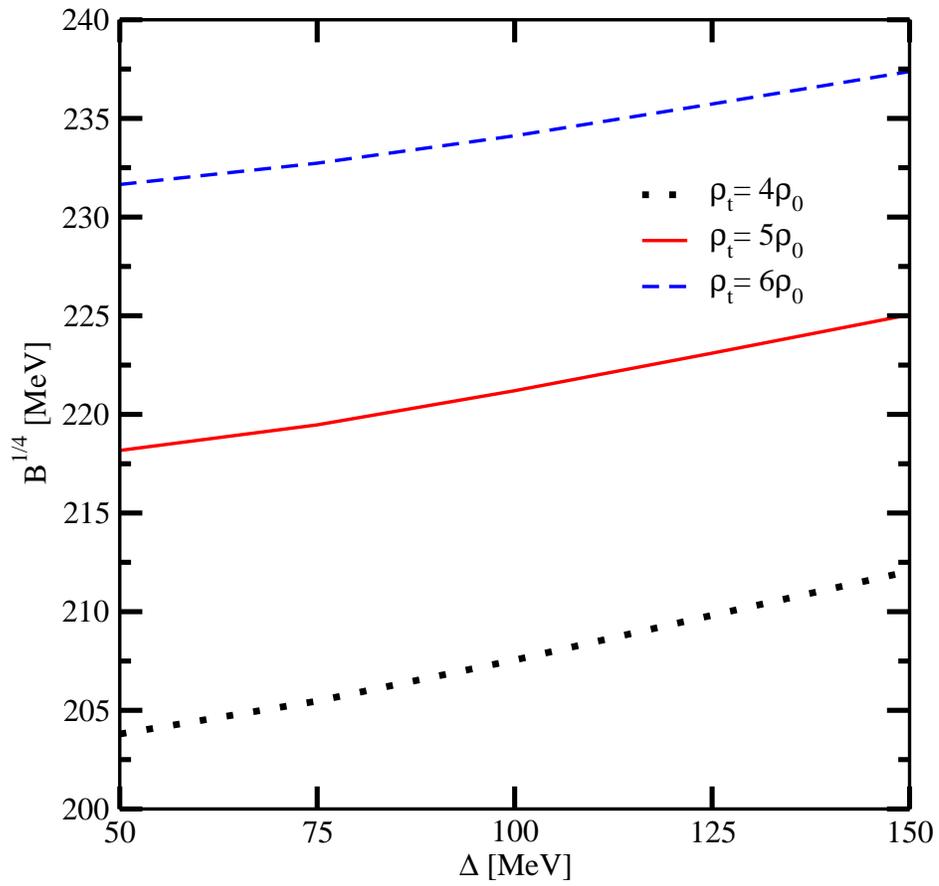}    
\caption{\label{fig:b14}(Color online)
Variations of bag constant as a function of the CFL gap parameter at
fixed values of hadron to CFL quark matter phase transition densities
$4\rho_0 - 6\rho_0$.  } 
 \end{figure}

\begin{figure}
\includegraphics[width=15cm]{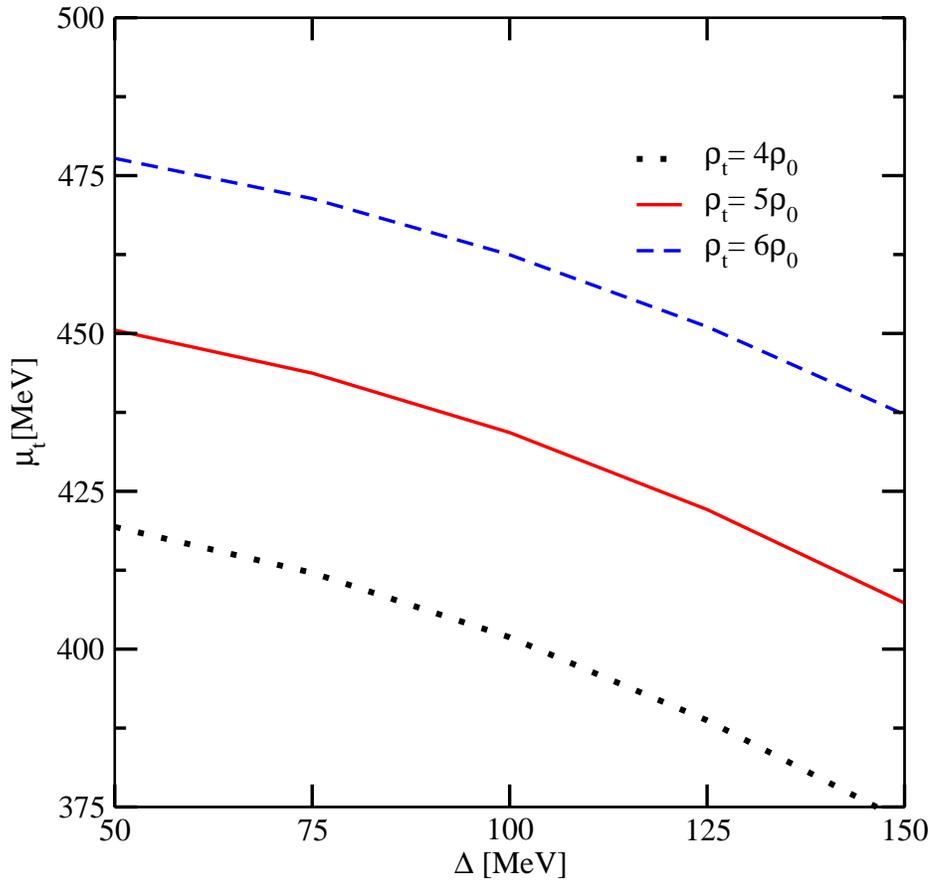}   
\caption{\label{fig:mu_del} (Color online)
Variations of the average value of quark chemical potential at the phase
transition density  as a function of the CFL gap parameter.  }
 \end{figure}

We study the properties of the hybrid stars, composed of CFL quark matter
core, for a set of EOSs obtained for different values of the CFL
gap parameter $\Delta$ and the deconfinement phase transition density
$\rho_t$. Instead of fixed values of  the bag constant as customarily
done \cite{Alford03,Banik03,Panda06}, we adjust the bag constant for each
values of the CFL gap parameter to yield the desired value of the phase
transition density.  We consider the values of the CFL gap parameter in
the range of $50 - 150\text{ MeV}$ as estimated and employed for the studies
of hybrid stars \cite{Alford98,Rapp98,Alford03,Alford05,Pagliara08}.
In Fig. \ref{fig:eos} we plot the several EOSs for the hybrid star matter
obtained for different values of the CFL gap parameter $\Delta$ with
$\rho_t= 4\rho_0 - 6\rho_0$.  The EOS for the hadron phase is obtained
within the framework of the extended FTRMF model as discussed in the
preceeding section.  In Ref. \cite{Dhiman07} we have obtained several
parameter sets for the extended FTRMF model for different values of
the coupling strength of the $\omega-$meson self-interaction term and
the neutron-skin thickness in $^{208}$Pb nucleus  as these are not well
determined from the presently available experimental data.  Each of
the parameterizations are consistent with  bulk properties of the
finite nuclei and nuclear matter. In the present work we have employed
the parameter set which corresponds to $\omega-$meson self-interaction
strength $\zeta = 0$ (Eq. \ref{eq:eden}) and neutron-skin thickness of
$0.2\text{ fm}$ in the $^{208}$Pb nucleus.  The choice of $\zeta = 0$ yields
stiff EOS for the hadronic matter at high density.  The EOSs of CFL
quark matter for different values of $\Delta$ and $\rho_t$ are obtained
using strange quark mass $m_s = 150\text{ MeV}$ and the constant $c = 0.3$ in
Eq. (\ref{eq:omega_quarks}).  In Fig. \ref{fig:b14} we plot the values of
the bag constant $B^{1/4}$ as a function of $\Delta$ for $\rho_t =
4\rho_0 - 6\rho_0$.  We get the values
of $B$ somewhat higher compared to the ones commonly used.  
In particular, the value of $B$ increases rapidly with the deconfinement
phase transition density.  This is because, stiffness of the EOS for
the hadronic matter as considered here increases rapidly with density.
Further, as the CFL gap parameter increases, the value of $B$ increases to
keep the phase transition density unaltered.  In Fig. \ref{fig:mu_del}
we plot the values of the average quark chemical potential $\mu_t$
at the deconfinement phase transition densities $\rho_t = 4\rho_0 -
6\rho_0$ as a function of the CFL gap parameter. For our choice of the
phase transition densities, the values of $\mu_t$ are in the range of
$375 - 500\text{ MeV}$ which are in reasonable agreement with the ones
obtained in Refs. \cite{Ippolito07,Ippolito08}.  The properties of the
static and rotating compact stars resulting from our set of EOSs are
computed using the code developed  by Stergioulas \cite{Stergioulas95}.

\subsection{Static sequences} 

\begin{figure}
\includegraphics[width=13.5cm]{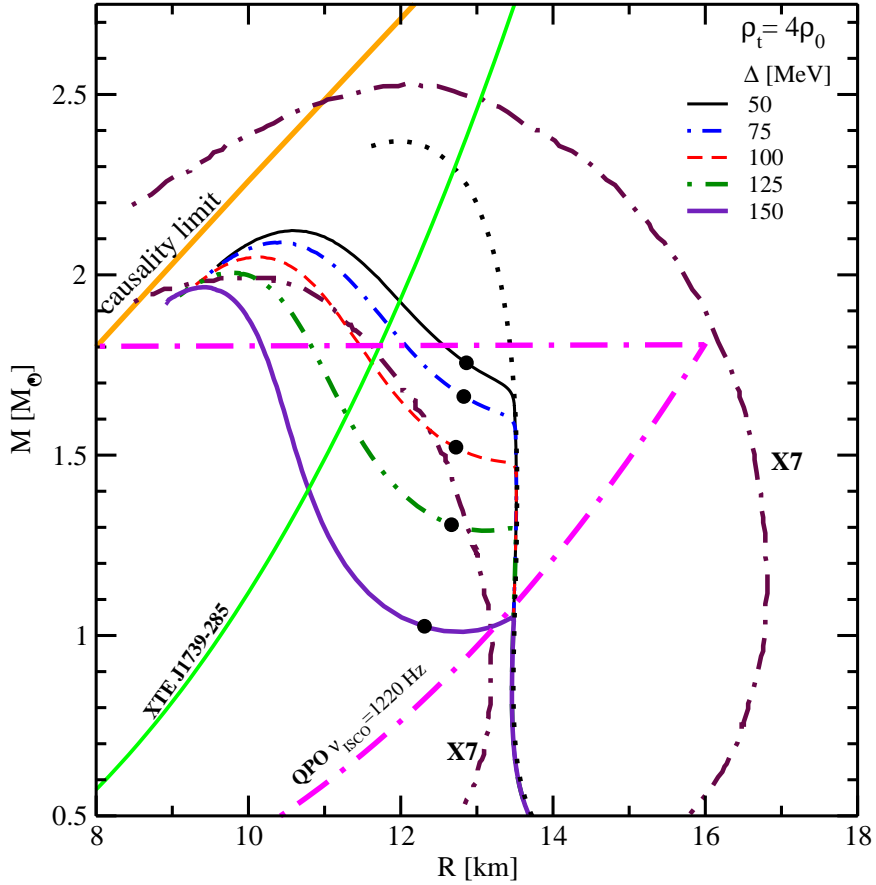}
\caption{\label{fig:mr_4} (Color online) 
Plots for the mass-radius relationships for the static sequences obtained
using fixed values of the CFL gap
parameter $\Delta $ ranging from $50 - 150\text{ MeV}$ with deconfinement phase
transition density $\rho_t = 4\rho_0$. The dotted black  line represent
the static sequence of compact stars composed of the hadronic matter.
The solid circle on each of the curves denote the end of the mixed phase.
The curves on the left to the solid circles represent the hybrid stars
with CFL core.  The dot-dashed maroon curves are the mass-radius contours
deduced with $90\%$ confidence  by fitting the high quality spectra from
the compact star X7 in the globular cluster 47 Tucanae \cite{Heinke06}.
}
\end{figure}

\begin{figure}
\begin{center}
\includegraphics[width=14cm]{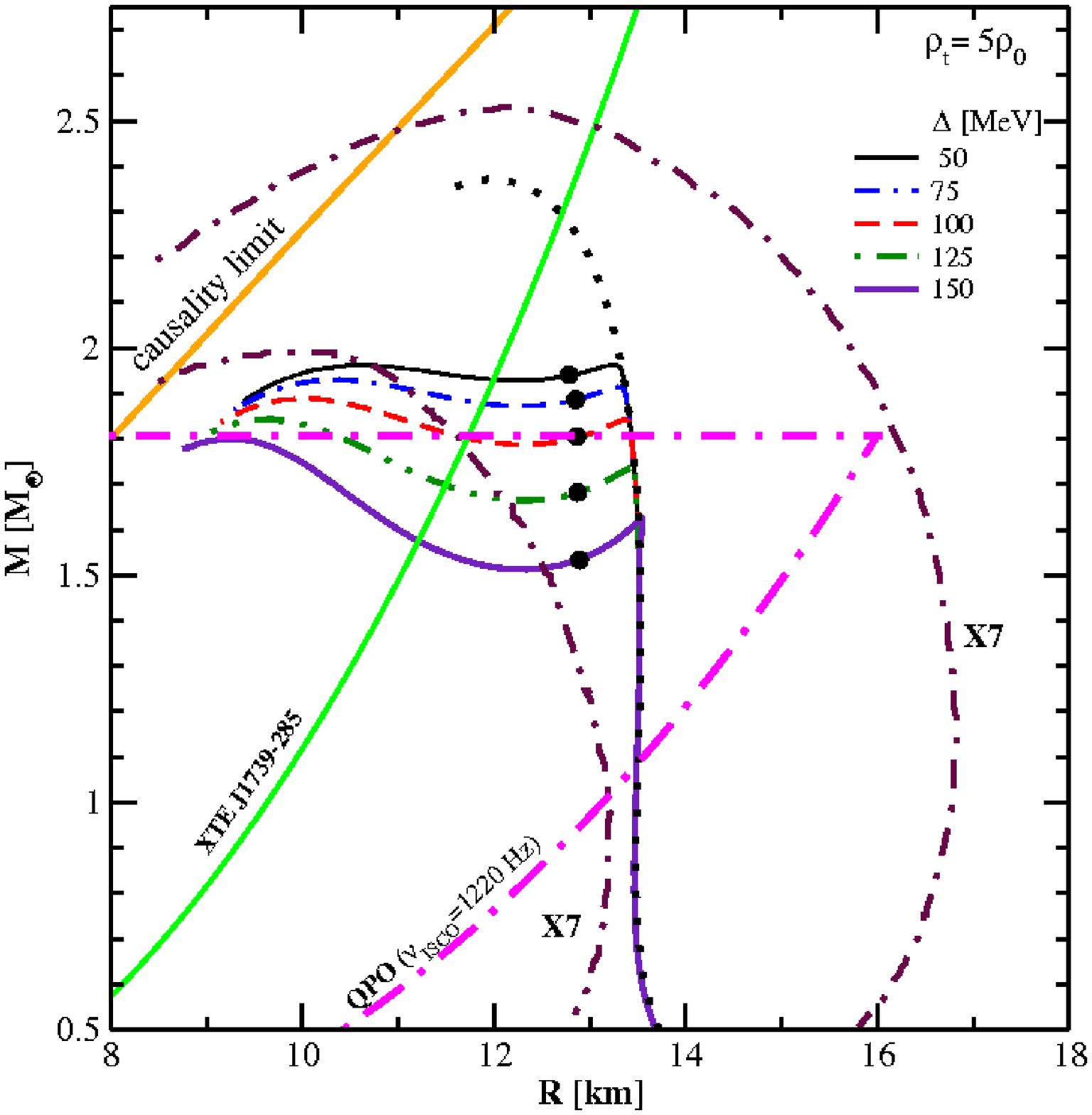}
\caption{\label{fig:mr_5} (Color online)
Same as Fig. \ref{fig:mr_4}, but, for the deconfinement phase transition
density $\rho_t=5\rho_0$.}
\end{center}
\end{figure}

\begin{figure}
\begin{center}
\includegraphics[width=14cm]{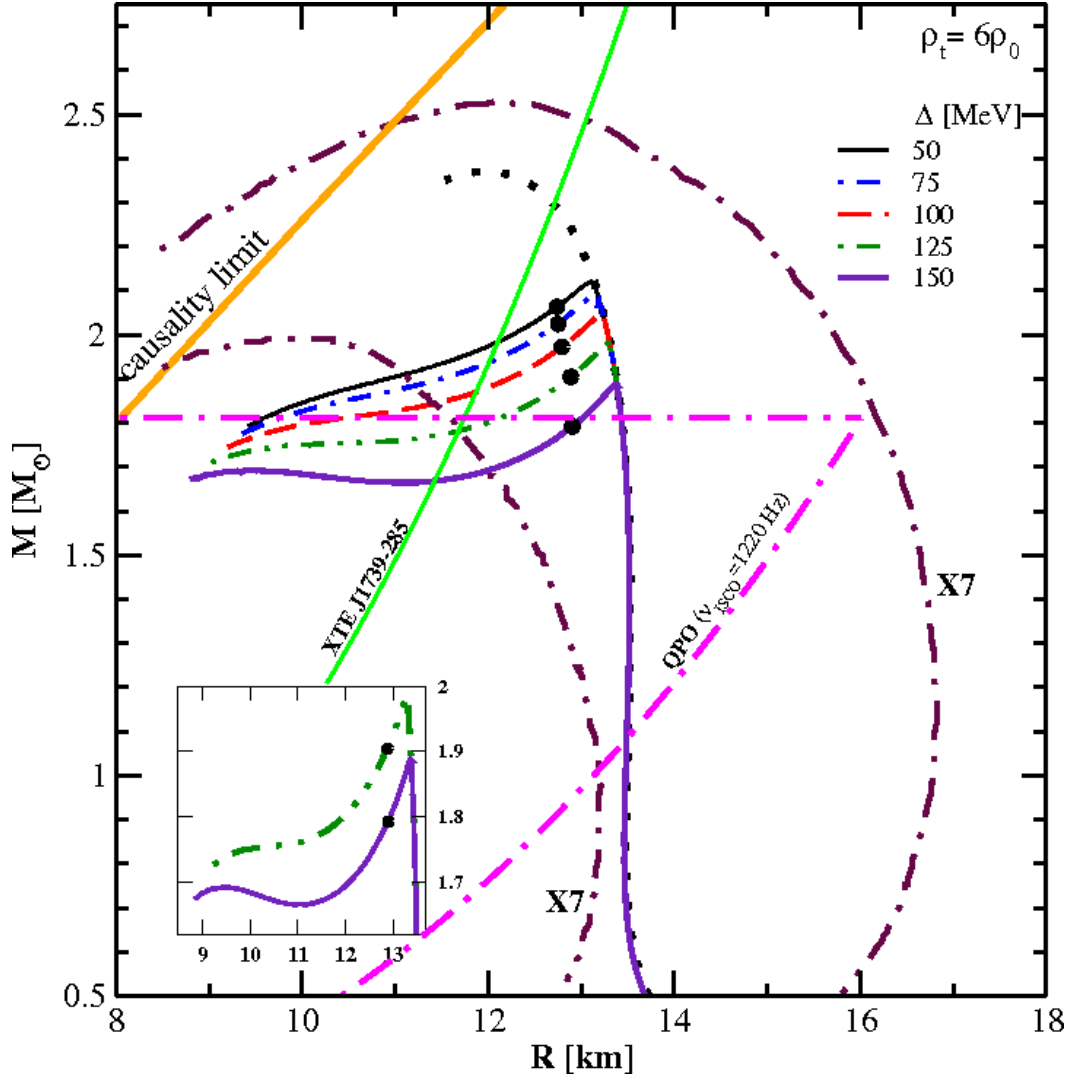}
\caption{\label{fig:mr_6} (Color online) 
Same as Fig. \ref{fig:mr_4}, but, for the deconfinement phase transition
density $\rho_t=6\rho_0$.
Inset highlights the appearance of third families of compact stars for the
CFL gap parameter $\Delta \geqslant 125\text{ MeV}$. }
\end{center}
\end{figure}

The sequence of static compact stars is obtained by varying the central
energy density $\epsilon_c$ for a given  EOS. For stable configuration,
\begin{equation}
 \frac{\partial M}{\partial \epsilon_c}  > 0
\label{eq:sbl}
\end{equation}
where, $M$ is the gravitational mass of the static 
compact star.  In Figs. \ref{fig:mr_4} -
\ref{fig:mr_6} we plot the mass-radius relationships for static sequences
obtained for various EOSs corresponding to the different values of the
CFL gap parameter $\Delta = 50 - 150\text{ MeV}$ with $\rho_t = 4\rho_0 -
6\rho_0$.  The solid circle on each of the curves marks the point at which
the deconfinement phase transition from hadron  to CFL quark matter occur.
The curves on the left to the solid circles represent the sequences of
the hybrid stars with CFL quark matter core.  The black dotted line
represents the static sequence of the compact stars composed of pure
hadronic matter.  The radius of hybrid stars decreases with increasing
central energy  density.  It can be seen  that the stable configurations
of hybrid stars with CFL quark matter core either bifurcate from the
hadronic sequence or form different branch so-called third family of
compact stars \cite{Glendenning00,Schaffner00}.  With $\rho_t = 4\rho_0 -
5\rho_0$, the stable configurations of the hybrid stars exist for all the
values of the CFL gap parameter considered.  In particular, for $\Delta
= 50 - 100\text{ MeV}$ with $\rho_t = 4\rho_0$, sequences of stable
configurations of the hybrid stars bifurcate  from the hadronic sequence
at  the central density exceeding the one at which the onset of mixed
phase occur.  For all the other cases with $\rho_t = 4\rho_0 - 5\rho_0$,
the hybrid stars belong to the third families of the compact stars.  When,
the value of $\rho_t$ is increased to $6\rho_0$, the hybrid stars with
CFL quark matter core become stable only for $\Delta \geqslant 125\text{
MeV}$.  We get the masses for such hybrid stars in the range of $1.0 -
2.1M_\odot$ with radii $9.3 - 13.5\text{km}$.

For the comparison, in Figs. \ref{fig:mr_4} - \ref{fig:mr_6} we plot
contours (dot-dashed/maroon) in the mass-radius plane which are deduced
with $90\%$ confidence, by fitting the high quality spectra from the
compact star X7 in the globular cluster 47 Tucanae \cite{Heinke06}. These
spectra were fitted within an improved hydrogen atmosphere model which
also  accounts for the variations in the surface gravity with mass and
radius of the compact stars.  These $M - R$ contours indicate that a
compact star with the canonical mass ($1.4M_\odot$)  should have radius
in the range of $\sim 13 - 17\text{ km}$, whereas, compact star with the
canonical radius ($10\text{ km}$) has mass $\sim 2 M_\odot$.  
The compact stars composed of only hadronic matter can satisfy either
the constraint on the radius at the canonical mass or the constraint on
the mass at the canonical radius \cite{Heinke06}.
Our results for the $\Delta = 50 - 75\text{
MeV}$ with $\rho_t = 4\rho_0 - 5\rho_0$ are bounded by the dot-dashed
maroon contours over broad range of mass and radius.  For these cases,
compact stars with canonical mass $1.4M_\odot$ have radii $13.5\text{km}$
and are composed of only hadronic matter. But, compact stars with radii
around the canonical value $10\text{km}$ have masses nearly $2M_\odot$
and are the hybrid stars with CFL quark matter core.  We also plot the $M
- R$ curves obtained using the constraints imposed by the discoveries of
the kHz quasi-periodic oscillations (QPOs) \cite{Miller98} and the X-ray
transient XTE J1739-285 \cite{Kaaret07}. The frequency of the innermost
stable circular orbit (ISCO)  inferred from the QPOs limits the  mass
of the non-rotating compact stars to be,
 \begin{equation}
M \leqslant \frac{2200\text{ Hz}}{\nu_{\scriptstyle ISCO}}M_\odot.
\end{equation} The values of $\nu_{\scriptstyle ISCO}$ are in the range
of $1220 - 1310\text{ Hz}$.  The compact star radius must be smaller
than the ISCO, which implies \cite{Miller98} \begin{equation} R \leqslant
\frac{19500\text{ Hz}}{\nu_{\scriptstyle ISCO}}\text{km}.  \end{equation}
Radii limits for  masses less than the upper limit scales with $M^{1/3}$.
The discovery of XTE J1739-285 suggests that it contains a compact star
rotating at 1122Hz. This, imposes the constraint on the maximum radius
of a non-rotating compact star with mass $M$ \cite{Lattimer07},
 \begin{equation}
\label{eq:1122} R_{\rm max} \leqslant 9.63\left (\frac{M}{M_\odot}\right
)^{1/3}.  \end{equation} We see that the hybrid stars with $M
\geqslant 1.4 M_\odot$ satisfy the constraint as expressed by
Eq. (\ref{eq:1122}). It is also found in Refs. \cite{Bejger07,Haensel08}
that mass of the compact star rotating with 1122 Hz is equal or larger
than $1.4M_\odot$.

In Table \ref{tab:static} we give the values of maximum masses
with corresponding  central energy densities and radii for the hadronic
and  the hybrid stars obtained using different values of the $\Delta$
and $\rho_t$.  It may be noted that, for a given $\Delta$ and $\rho_t$,
the central energy density for the maximum mass of the hadronic star
corresponds to the one at which the onset of mixed phase occurs (see
also Fig.  \ref{fig:eos}).  The values of maximum mass $1.7 - 2.1M_\odot$
for the hybrid stars  are consistent with the currently measured maximum
mass 1.76$\pm0.20M_\odot$  of PSR J0437-4715 \cite{Verbiest08} obtained
by the precise determination of the orbital inclination angle.  We note,
for $\rho_t = 5\rho_0$, maximum mass of the hybrid stars with CFL core
are nearly equal to their second family counterpart composed of hadrons.
The hybrid stars with CFL core are smaller by about $30\%$ compared
to those second family counterpart.  Thus, the hybrid stars with CFL
core are expected to rotate significantly faster in comparison to the
hadronic stars.  We also remark that our results for the mass-radius
relationship are somewhat similar to the ones obtained using the NJL
like model \cite{Ippolito08,Pagliara08}, for the hybrid stars composed of
CFL or CCS quark matter core.

\subsection{Mass-shedding limit sequences }
\begin{figure}
\begin{center}
\includegraphics[width=14cm]{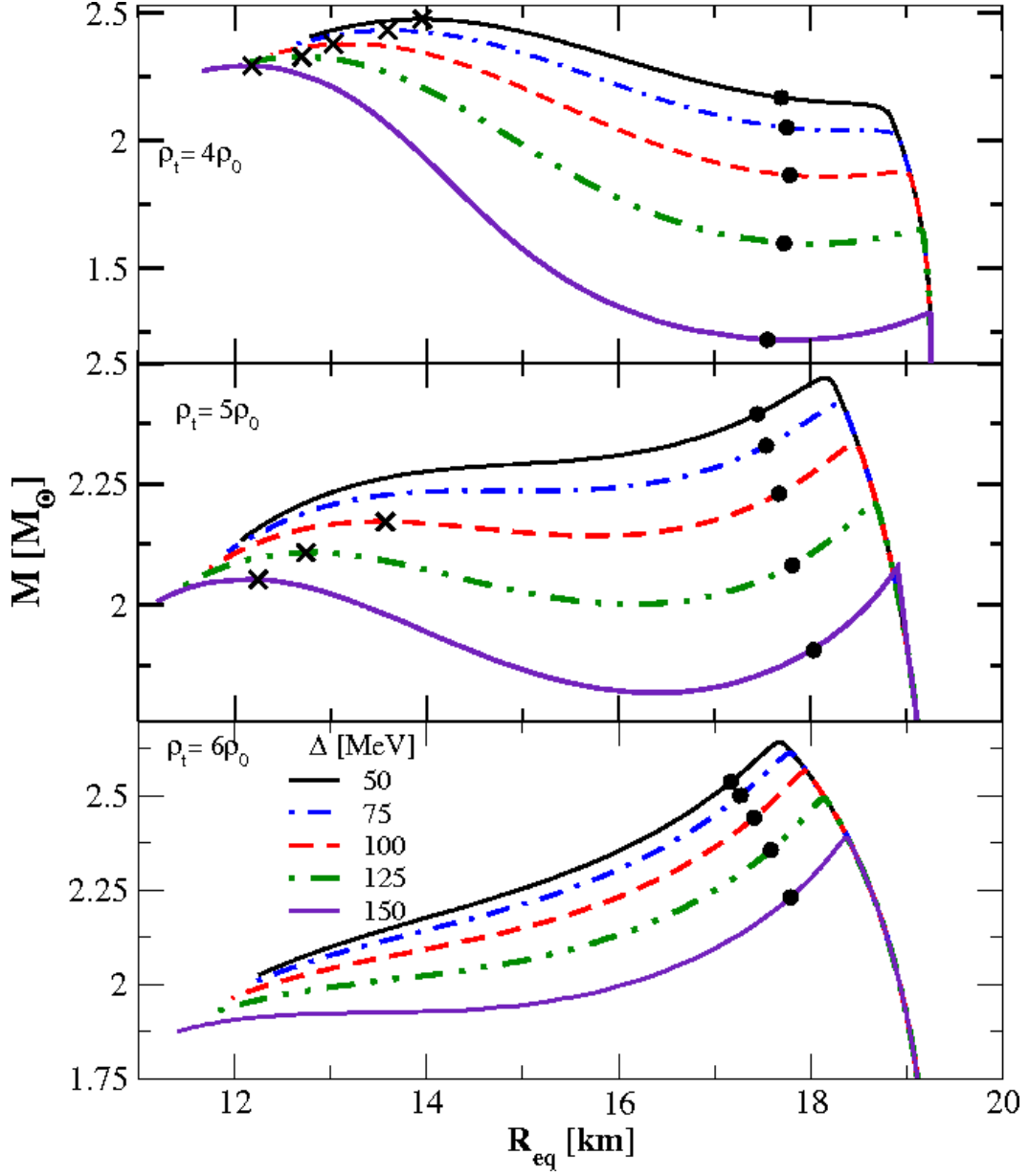}
\caption{\label{fig:kep_456} (Color online) 
Relationship between mass M and the circumferential equatorial radius
$R_{\text{eq}}$  for mass-shedding limit  sequences 
for different values of the CFL gap parameter  and the phase transition
densities $4\rho_0$ (upper panel), $5\rho_0$ (middle panel) and $6\rho_0$
(lower panel). The curves on the left to the  solid circles represent the 
hybrid stars with CFL core. The symbol cross
$(\times)$ on the different curves marks the maximum mass of the hybrid
star.}

\end{center}
\end{figure}

In Fig. \ref{fig:kep_456}  we plot relationship between mass  and
the circumferential equatorial radius $R_{\text{eq}}$ for the
mass-shedding limit sequences obtained for the EOSs corresponding to
different values of $\Delta$ and $\rho_t$.  The portion of the curves
left to  the solid circles represent the hybrid stars with CFL core.
The symbol cross $(\times)$ on the different curves marks the maximum
mass of the hybrid star.  For $\rho_t = 5\rho_0$ with  $\Delta = 50$
and 75 MeV, third family of compact stars with CFL core disappears
from the mass-shedding limit sequence, though, they exist in the static
sequence as can be seen from Fig. \ref{fig:mr_5}. Similar situation is
encountered for $\Delta \geqslant 125\text{ MeV}$ with $\rho_t = 6\rho_0$
(see also Fig. \ref{fig:mr_6}).  In Table \ref{tab:kepler}, we give the
values of the central energy density, radius and Kepler (mass-shedding)
frequency $f_K$  at the maximum mass for the hadronic and hybrid stars
corresponding to different $\Delta$ and $\rho_t$.  The values of the
Keplerian frequency at the maximum hybrid star mass is in the range of
$1.7 - 2$ kHz, whereas, the Keplerian frequency at the maximum mass for
the hadronic stars is $\sim 1$ kHz.

\begin{figure}
\includegraphics[width=14cm]{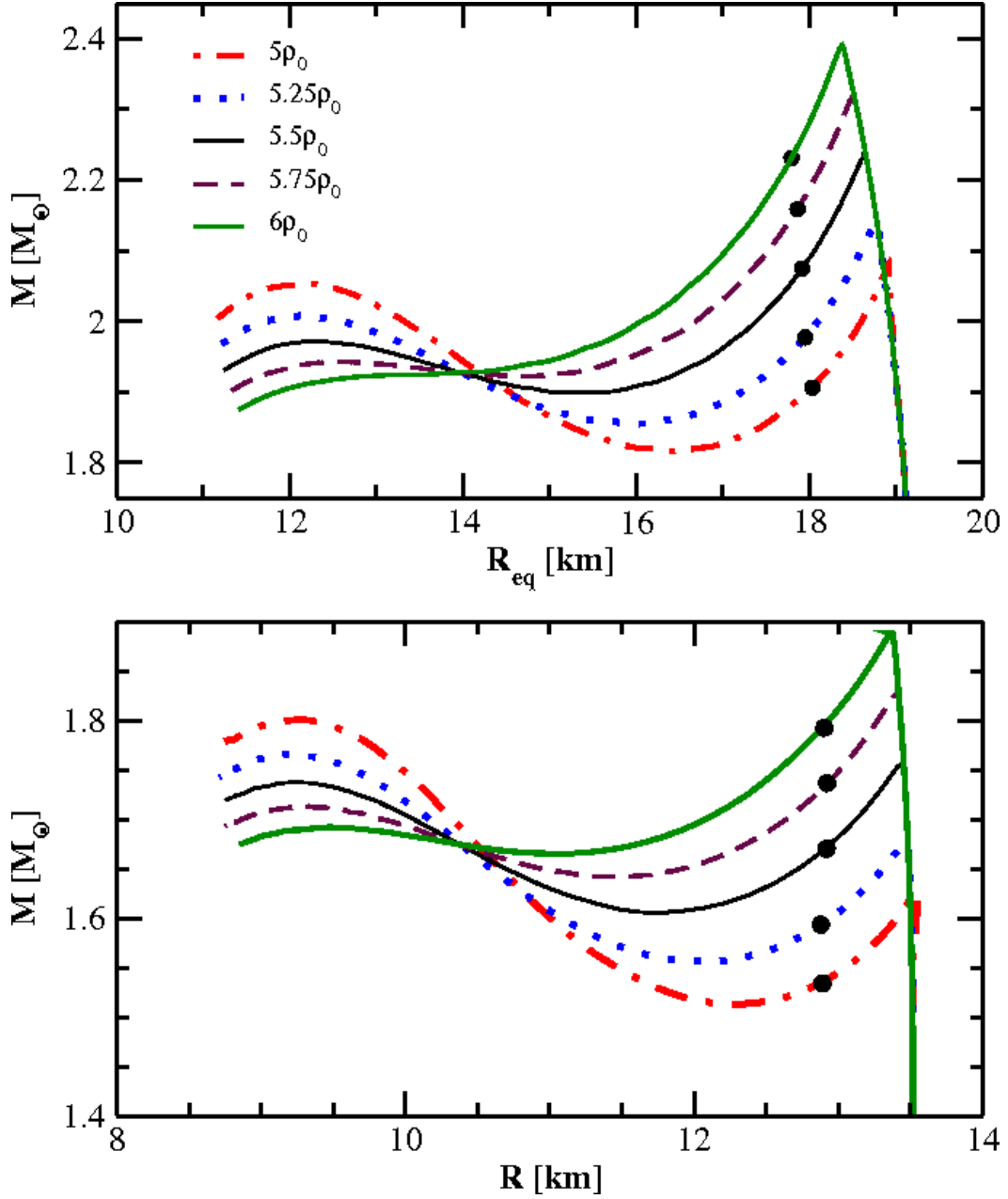}
\caption{\label{fig:kep_del150} (Color online) 
Plots for the mass-shedding limit sequences (upper panel) and the static
sequences (lower panel) for the CFL gap parameter $\Delta = 150\text{ MeV}$ with
deconfinement phase transition density $\rho_t = 5\rho_0 - 6\rho_0$.
The parts of the curves left to the solid circles in the upper
and lower panel represent the sequences of hybrid stars with the CFL
quark matter core.
}
\end{figure}

We now examine several cases for which third family  of compact stars
with CFL core  appears in the  static sequence, but, disappear from
the mass-shedding limit sequence.  We observe that for these cases,
maximum mass of the third family compact stars is lower than their second
family counterpart.  In Fig. \ref{fig:kep_del150}, we have plotted the
mass-shedding limit sequences (upper panel) and the static sequences
(lower panel) for $\Delta = 150\text{ MeV}$ with $\rho_t$ ranging from
$5\rho_0 - 6\rho_0$.  We clearly see that the third family of compact
stars tend to disappear beyond $\rho_t > 5.5\rho_0$. Strikingly, at
$\rho_t = 5.5\rho_0$, maximum masses of the second and third family
compact stars are nearly equal in the static limit.  We find similar
outcome for  the other cases (not shown here).  Thus, it seems there
exist a critical value of $\rho_t$ for a given $\Delta$ beyond which
third family of compact stars with CFL core  tend  to disappear with
increase in $\rho_t$.  Below the critical value of $\rho_t$, such
hybrid stars in the non-rotating limit have maximum mass higher than
their counterpart composed of hadronic matter.  Our these results are
substantiated by the earlier calculations performed using different
EOSs \cite{Ippolito08,Banik05,Bhattacharyya05}.  The EOS used in
Ref. \cite{Ippolito08} yields third family of compact stars in the
static as well as in the mass-shedding limit sequences. For this EOS,
the maximum mass of the compact star belonging to third family is larger
by about $0.1M_\odot$ compared to its second family counterpart.  On the
other hand, for the EOSs used in Refs.  \cite{Banik05,Bhattacharyya05},
the third family compact stars exist in the static limit and disappears
from the mass-shedding limit sequences. For these EOSs, maximum mass
of the second and third family compact stars are nearly equal.  In Ref.
\cite{Banik05}, the maximum masses for the second and third family compact
stars are found to be $1.57M_\odot$ and $1.55M_\odot$, respectively.
In Ref.  \cite{Bhattacharyya05}, the maximum  masses for the second
and third family compact stars are $1.36M_\odot$ and $1.  38M_\odot$,
respectively.

\subsection{Critical rotation frequency} 
\begin{figure}
\includegraphics[width=14cm]{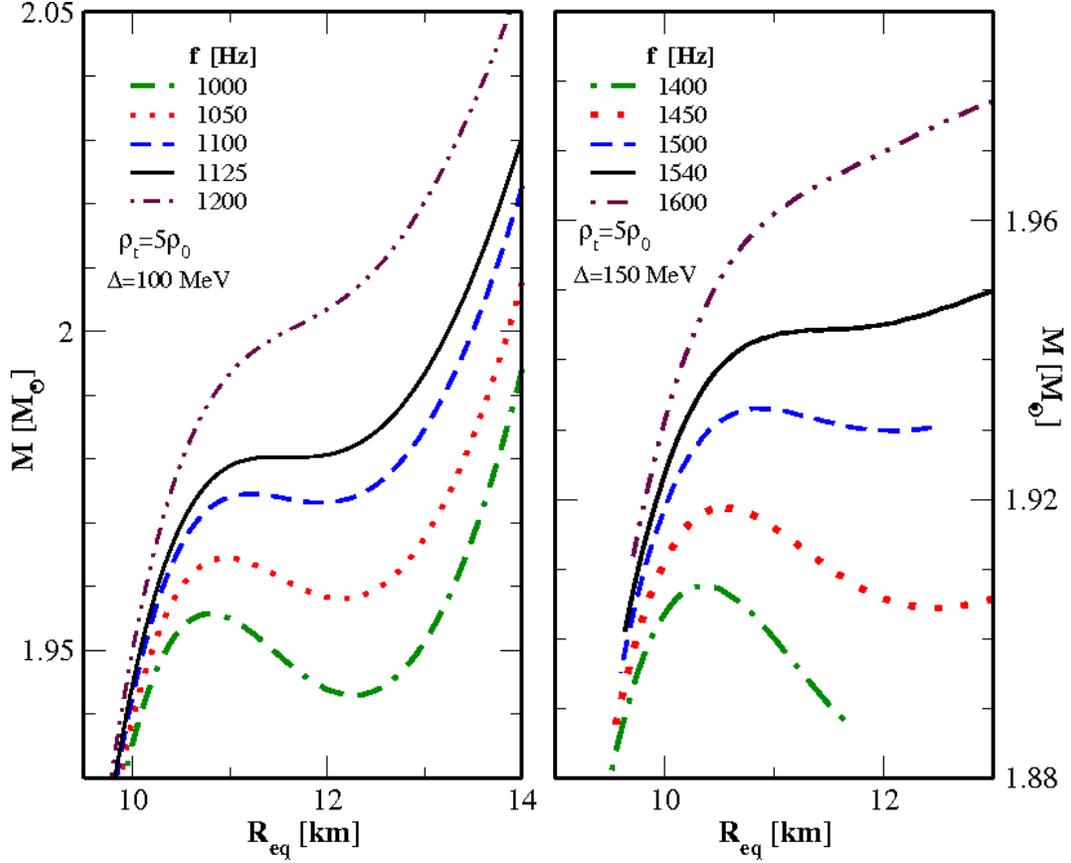}
\caption{\label{fig:omega_crit} (Color online) 
Plots for the mass verses circumferential equatorial radius $R_{\rm eq}$
at fixed values of the rotational frequency. The black solid lines
represent the results obtained at the critical frequencies $f_{\rm
crit}$.  For $f>f_{\rm crit}$, third families of compact stars
do not exist.
The values of $\rho_t$ and $\Delta$ considered are such that they yield
third family compact stars in the static as well as in the mass-shedding
limit sequences. 
}
\end{figure}

\begin{figure}
\includegraphics[width=14cm]{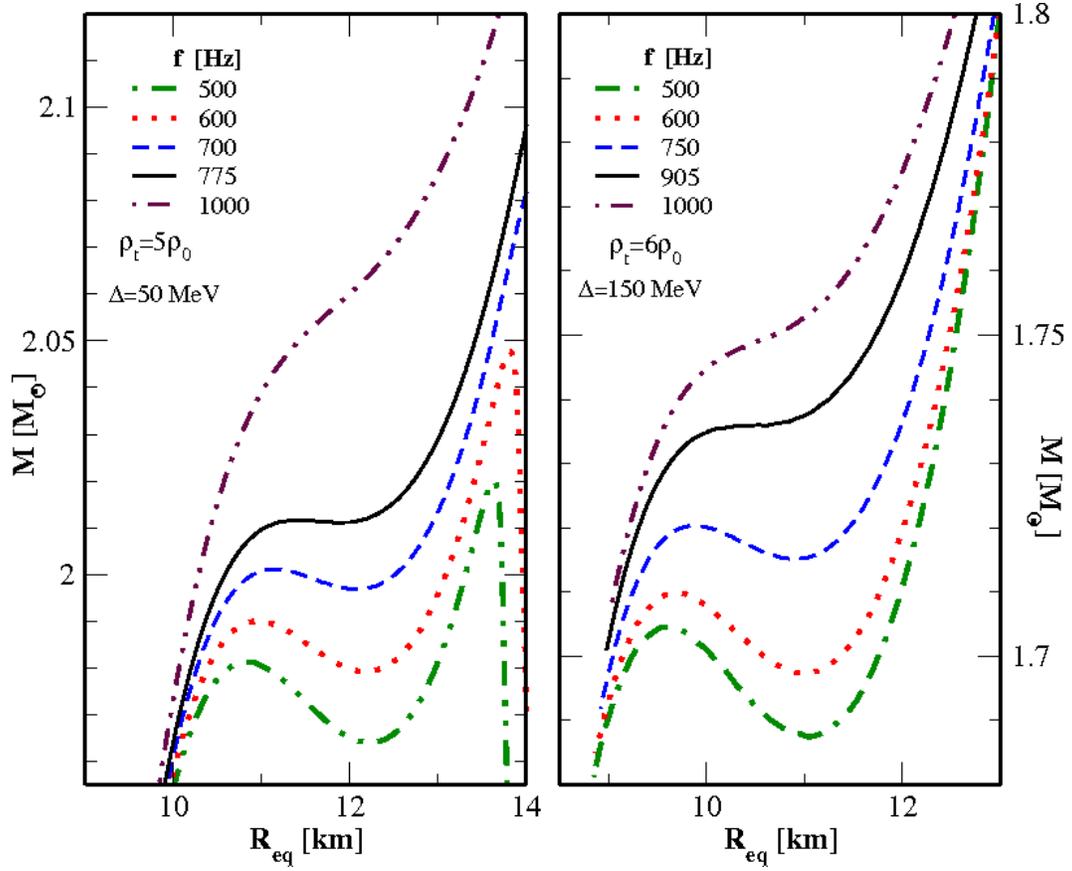}
\caption{\label{fig:omega_crit1} (Color online) 
Same as Fig. \ref{fig:omega_crit}. But,  
the values of $\rho_t$ and $\Delta$ considered are such that they yield
third family compact stars in the static sequences which disappears from
the mass-shedding limit sequences.
}
\end{figure}

We have computed the values of the  maximum or the critical rotation
frequency $f_{\rm crit}$  for the stable configurations of  hybrid stars
with CFL core. The stable configurations of the compact stars rotating at a
given frequency $f$ satisfy,
\begin{equation}
\left ( \frac{\partial M}{\partial \epsilon_c}\right )_f  > 0.
\label{eq:stbl1}
\end{equation}
The Eq. (\ref{eq:stbl1}) is satisfied only for  $f \leqslant f_{\rm
crit}$.  To locate the critical frequency we first obtained the variation
in the mass as a function of $\epsilon_c$ at fixed frequencies in
steps of $50$Hz.  Then, for appropriate interval of the frequency, the
calculations were repeated by varying the frequency in steps of $5$Hz to
determine the value of $f_{\rm crit}$.  In Figs. \ref{fig:omega_crit}
and \ref{fig:omega_crit1} we plot the $M - R_{\rm eq}$ curves at
fixed values of the rotational frequency.  
The black solid lines represent the results obtained at the
$f = f_{\rm crit}$.  
For the clarity, we mainly focus on the regions of  the $M - R_{eq}$
curves corresponding to the sequences of the hybrid stars which  are
relevant in the present context.  In this region, the value of $R_{\rm
eq}$ decreases with increase in $\epsilon_c$.  The results presented in
Fig. \ref{fig:omega_crit} correspond to the cases for which third family
of compact stars exists in the static as well as in the  mass-shedding
limit sequences. In Fig. \ref{fig:omega_crit1}, we consider the cases
for which third family of compact stars exist in the static limit, but,
disappears from the mass-shedding limit sequences.  We see that the value
of $f_{\rm crit}$, for the cases presented in Fig. \ref{fig:omega_crit},
are larger than the highest observed rotation frequency  1122 Hz.
The value of $f_{\rm crit} \simeq 775 - 900\text{ Hz} $ for the cases
presented in Fig. \ref{fig:omega_crit1}.  For $\rho_t = 4\rho_0$, the
value of $f_{\rm crit}$ increases from $1370 - 1805$ Hz as the CFL gap
parameter $\Delta $ increases from 50 MeV to 150 MeV.  It may be pointed
out that the situation analogous  to that of  Fig. \ref{fig:omega_crit1}
is encountered in Refs.  \cite{Banik05,Bhattacharyya05}, but, the value
of  $f_{\rm crit}$ is about $350 - 650\text{ Hz}$.

In Table \ref{tab:omega_crit}, we summarize the properties  at the
maximum mass of the hybrid stars with CFL core rotating with critical
frequency.  We compare the values of $\epsilon_c(f_{\rm crit})$ as
given in this table with the maximum ($\epsilon_c(0)$) and the minimum
( $\epsilon_c^\prime(0)$) energy densities at the centre of stable
configurations of non-rotating hybrid  stars. In the last two columns of
Table \ref{tab:omega_crit} we give the values of $\partial \epsilon_1$
and $\partial \epsilon_2$ calculated as,

 \begin{equation}
\partial \epsilon_1 = 1 -\frac {\epsilon_c(f_{\rm crit})} {\epsilon_c(0)},
\label{eq:dele1}
\end{equation}
and
\begin{equation}
\partial \epsilon_2 = 1 -\frac {\epsilon_c^\prime(0)} {\epsilon_c(0)}.
\label{eq:dele2}
\end{equation}
The values of $\epsilon_c(0)$ and $\epsilon_c(f_{\rm crit})$
are taken from 6th and 4th columns of Tables \ref{tab:static} and
\ref{tab:omega_crit}, respectively.  It is clear from the values of
$\partial \epsilon_1$ that the central energy density $\epsilon_c(f_{\rm
crit})$ is smaller than $\epsilon_c(0)$ by $25 - 30 \%$.  The values of
$\partial \epsilon_2$  are noticeably larger for the cases considered in
Fig. \ref{fig:omega_crit} than those of Fig.  \ref{fig:omega_crit1}. For
$\rho_t = 4\rho_0$ with $\Delta = 150$ MeV, we get $\partial \epsilon_1 =
0.25$ and $\partial \epsilon_2 = 0.62$.  We would like to add that the EOS
used in Ref. \cite{Bhattacharyya05} yields $\partial \epsilon_2\simeq 0.4$
for which  third family of compact stars disappears from the mass-shedding
limit sequence, but, exists in the static limit.  It thus appears that
$\partial \epsilon_2 \lesssim 0.4$  disfavours the appearance of third
family of compact stars in the mass-shedding limit sequence even if it
exists in the static limit.
It may be noted that the values of $f_{\rm crit}$  (Table
\ref{tab:omega_crit}) is lower than the Kepler
frequency (Table \ref{tab:kepler}) for the cases considered in Fig.
\ref{fig:omega_crit}. We would like to point out that the vlaue of
$f_{\rm crit}$ for a given EOS represents maximum rotation frequency
for which the stability condition as given by Eq. (\ref{eq:stbl1})
is satisfied.  Whereas, the mass-shedding limit sequences can not be
subjected to Eq. (\ref{eq:stbl1}).  Along these sequences, rotation
frequency corresponds to the Kepler frequency which increases with the
central energy density.  This leads to the values of Kepler frequency
at maximum mass higher than the $f_{crit}$ for a given EOS.

\section{Conclusions} 
We construct static and mass-shedding limit sequences of hybrid stars
for a set of EOSs obtained for different values of the CFL gap parameter
and the deconfinement phase transition density.  The hybrid stars
considered are  composed of CFL quark matter at the core, nuclear matter
at the crust and mixed phase in the intermediate region.  The hadronic
part of the EOS is obtained  using appropriately calibrated extended
field theoretical based relativistic mean-field model.  The EOSs of quark
matter in the CFL phase corresponding to different values of the CFL gap
parameter and the deconfinement phase transition density    are obtained
using MIT bag model with an additional parameter that mimics the effect
of including perturbative QCD corrections.  The CFL gap parameter ranges
from $50 - 150\text{ MeV}$ with the deconfinement phase transition
density ranging from $4\rho_0 - 6\rho_0$ ($\rho_0 = 0.16\text{ fm}^{-3}$).

We find the existence of the stable configurations of the static hybrid
stars for all the different values of the CFL gap parameter considered with the
deconfinement phase transition density  $4\rho_0 - 5\rho_0$.  For the
cases with CFL gap parameters $50 - 100\text{ MeV}$ with  deconfinement phase
transition density  $4\rho_0$, the sequences of stable configurations of
hybrid stars bifurcate from hadronic sequence when the central density
exceeds the one at which the onset of mixed phase occur.  In all the other
cases, the  stable configurations of hybrid stars form the third family
of compact stars.  When the deconfinement phase transition density is
increased to $6\rho_0$ the stable configurations of hybrid stars exist
only for CFL gap parameter $\Delta \geqslant 125\text{ MeV}$. For the CFL gap
parameter $50 - 75\text{ MeV}$ with the deconfinement phase transition densities
with $4\rho_0 - 5\rho_0$, the mass-radius relationship over broad range
of mass and radius are in harmony with those deduced by applying improved
hydrogen atmosphere model to fit the high quality spectra from compact
star X7 in the globular cluster 47 Tucanae.  The values of maximum
mass $1.7 - 2.1M_\odot$ for the hybrid stars  are consistent with the
currently measured maximum mass 1.76$\pm0.20M_\odot$  of PSR J0437-4715
\cite{Verbiest08}.

We find for several cases that the third family of compact stars
disappear from the mass-shedding limit sequences, though, they appear in
the corresponding static sequences.  Our investigation suggest that the
third family compact stars is more likely to  appear in the mass-shedding
limit sequence provided they have maximum mass in the static limit higher
than their counterpart composed of pure hadronic matter.  Further, we
have calculated the quantity $\partial \epsilon_2$ (Eq.  \ref{eq:dele2})
obtained using the minimum and the maximum values of the central energy
densities  for the stable configurations of the static hybrid stars. The
values of $\partial \epsilon_2 $ is less than $ 0.4$ for the cases for
which third family of the compact stars disappears from the mass-shedding
limit sequences, but, exists in the static limit. Except for these cases,
the values of the critical rotation  frequency for the hybrid stars
with CFL core are larger than the highest observed frequency 1122 Hz.
The relationship between the values of $\delta\epsilon_2$ and the
disappearance of the third family of compact stars from the mass-shedding
limit sequences as observed in the present work is only preliminary.
To establish this relationship, more investigations must be carried out
using wide variety of EOSs.


\newpage

\begin{table}
\caption{\label{tab:static}
The maximum mass of hybrid stars with CFL core and hadron stars in
the static limit  and corresponding central energy density and radius
obtained for different values of the CFL gap parameter $\Delta$ and the
deconfinement phase transition density $\rho_t$.
 }

\begin{tabular}{cccccccc}
\hline
& & &Hadron Stars& &Hybrid Stars&\\
\hline
$\rho_t$&$\Delta$&$ \epsilon$ &M &R &$\epsilon$ &M &R\\
&(Mev) &$(10^{15}$g/cm$^3)$&$M_\odot$ &(km)&$(10^{15}$g/cm$^3)$&$M_\odot$&(km)\\
\hline
&50& 0.721&1.63& 13.49&2.716&2.12&10.57\\
4$\rho_0$&100&0.642&1.45&13.52&2.863&2.05&10.16\\
&150&0.512&1.04&13.48&3.244&1.97&9.42\\
\hline
&50&0.873&1.94&13.34&2.863&1.96&10.53\\
$5\rho_0$&100&0.808&1.83&13.42&3.072&1.89&10.10\\
&150&0.690&1.62&13.55&3.663&1.80&9.28\\
\hline
&50&1.032&2.10&13.15& -& -& -\\
$6\rho_0$&100&0.959&2.04&13.24&-&-&-\\
&150&0.850&1.88&13.38&4.071&1.69&9.30\\
\hline
\end{tabular}
\end{table}

\newpage
\begin{table}
\caption{\label{tab:kepler}
The central energy density, radius and Kepler frequency  at the maximum
mass for the hybrid stars  with CFL core and hadron stars  obtained for
different values of the CFL gap parameter $\Delta$ and the deconfinement
phase transition density $\rho_t$.  }

\vskip 1cm
\begin{tabular}{cccccccccc}
\hline
& & &&Hadron Stars& &&Hybrid Stars&\\
\hline
$\rho_t$&$\Delta$&$ \epsilon$ &M &R &f$_K$&$\epsilon$ &M &R&f$_K$\\
&(Mev) &$(10^{15}$g/cm$^3)$& $M_\odot$&(km)&(Hz)& $(10^{15}$g/cm$^3)$&$M_\odot$&(km)&(Hz)\\
\hline
&50&0.721&2.09&18.84&1028&2.383 &2.48&13.95&1698\\
$4\rho_0$&100&0.642&1.84&19.06&953&2.665&2.38&13.21&1803\\
&150&0.512&1.32&19.24&801&2.934&2.29&12.18&1993\\ 
\hline
&50&0.873 &2.45&18.25&1158&- &-&-&-\\
$5\rho_0$&100&0.808 &2.32&18.51&1107&2.842 &2.17&13.44&1687\\
&150&0.690 & 2.08&18.95&1019&3.339 & 2.05&12.20&1889\\
\hline
\end{tabular}
\end{table}
\newpage
\begin{table}
\caption{\label{tab:omega_crit}
Properties at the maximum  mass of hybrid stars with CFL core
rotating at the critical frequency 
as obtained for different values of the CFL gap parameter $\Delta$ and the
deconfinement phase transition density $\rho_t$.  The values of
$\partial\epsilon_1$ and $\partial\epsilon_2$  are obtained using
Eqs. (\ref{eq:dele1}) and (\ref{eq:dele2}).
 }

\begin{tabular}{cccccccc}
\hline
$\rho_t$&$\Delta$& $f_{\rm crit}$&$ \epsilon$ &M &$R_{\rm eq}$ &
$\partial \epsilon_1$  & $\partial \epsilon_2$\\
&(Mev) &(Hz)& $(10^{15}$g/cm$^3)$&$(M_\odot)$ &(km)& \\
\hline
&50& 775& 2.160& 2.01& 11.58& 0.25& 0.35\\
$5\rho_0$&100& 1125& 2.205& 1.98& 11.49& 0.28& 0.45\\
&150&1540& 2.685& 1.94& 11.30& 0.27& 0.55\\
\hline
$6\rho_0$&150&905& 2.951& 1.74& 10.62& 0.28& 0.30\\
\hline
\end{tabular}
\end{table}

\vspace*{4 in}
\newpage


\end{document}